\documentclass{ECOC-author}

\begin{document}

\title{FIRST TRANSMISSION OF A 12D FORMAT ACROSS 3 COUPLED SPATIAL MODES OF A 3-CORE COUPLED-CORE FIBER AT A SPECTRAL EFFICIENCY OF 4~BITS/S/HZ}

\author{Ren\'e-Jean Essiambre\ad{ 1}, \corr\, Roland Ryf\ad{ 1}, Sjoerd van der Heide\ad{ 1,2}, Juan I.     Bonetti\ad{ 1,3}, Hanzi Huang\ad{ 1,4}, Murali Kodialam\ad{ 1}, Francisco Javier Garc\'ia-G\'omez\ad{ 1,5}, Ellsworth C. Burrows\ad{ 1}, Juan C. Alvarado-Zacarias\ad{ 1,6}, Rodrigo Amezcua-Correa\ad{ 6}, Xi Chen\ad{ 1}, Nicolas K. Fontaine\ad{ 1} and  Haoshuo Chen\ad{ 1}}
% Roland Ryf\ad{1}, Corresponding author\ad{3}\corr\ (use style: unabbreviated first name, middle name initials, last name e.g. John A K Smith\ad{1}, Edward Jones\ad{2})}
\address{\add{1}{Nokia Bell Laboratories, Holmdel, NJ, USA, 07733}
\add{2}{Department of Electrical Engineering, Eindhoven Univ. of Technology, The Netherlands.}
\add{3}{Grupo de Comunicaciones \'Opticas, Instituto Balseiro, Av. Bustillo km 9.500, Bariloche (R8402AGP), Argentina}
% Consejo Nacional de Investigaciones Científicas y Técnicas (CONICET), Argentina
\add{4}{Key laboratory of Specialty Fiber Optics and Optical Access Networks, Shanghai Univ., 200444 Shanghai, China}
\add{5}{Institute for Commun. Engineering, Technical Univ. of Munich, Theresienstrasse 90, 80333, M\"unchen, Germany }
\add{6}{CREOL, The University of Central Florida, Orlando, Florida 32816, USA }
\email{rene.essiambre@nokia.com}}

% \keywords{Enter a maximum of \underline{five} keywords here (use style: KEYWORD ONE, KEYWORD TWO, \ldots)}

\begin{abstract}
We demonstrate the first transmission of a new twelve-dimensional modulation format over a three-core coupled-core multicore fiber. The format occupies a single time slot spread across all three linearly-coupled spatial modes and shows improved MI and GMI after transmission compared to PDM-QPSK.
\end{abstract}

\maketitle

\section{Introduction}

Advanced modulation formats are essential to achieve high spectral efficiencies (SEs) in the presence of additive white Gaussian noise (AWGN). Various bidimensional (2D) modulation formats with different SEs have been considered to minimize bit-error rate (BER), maximize mutual information (MI) or generalized mutual information (GMI) for a given constellation and signal-to-noise ratio (SNR)~\cite{ess2010a}. There are several ways to optimize constellations over a given channel, including geometric~\cite{lot2013}, probabilistic~\cite{boc2015,buc2016,cho2017,gha2017} and hybrid shapings~\cite{cai2017}.

In fiber-optic communication systems, the presence of nonlinear effects can play a major role in determining system performance. Some formats show natural resistance to nonlinear distortions like polarisation-division multiplexed quaternary phase shift keying (PDM-QPSK) being both constant power and on a cartesian grid (suitable for Gray mapping). Among advanced formats geared towards mitigating nonlinear effects, one notes formats resistant to phase distortions~\cite{pfa2011}, shaped using shell mapping~\cite{gel2016}, 4D using points symmetrically located on two rings~\cite{koj2014,koj2017,hei2019}, 8D constellations based on selecting points within PDM-QPSK~\cite{shi2014} or optimized for GMI~\cite{che2019}, 12D~\cite{bul2011} and 16D~\cite{rad2015}, the last two having large minimum distances but limited to low SE. So far, these formats have been implemented as time sequences of 4D symbols or across uncoupled fiber core on a single time slot.
% A format particularly efficient for fiber-optic communication systems is the PDM-QPSK format as it has constant power and is Gray mapped.
% forcing the use of additional techniques of digital signal processing for nonlinearity mitigation~\cite{car2017}.
% NOT CITED
% Not cited yet
% 4D~\cite{hei19}
% space-division multiplexing (SDM) on uncoupled-core multicore fiber~\cite{put14}
% Ericksson in ecoc 2013
% \cite{koj2014,koi2016,koj2017}
% \cite{gha2016}

In this paper, we introduce a new 12D modulation format, implemented across the three spatial modes over a single time slot. We show improved transmission performance compared to PDM-QPSK at the same SE. To our knowledge, this is the first experimental transmission of a multidimensional constellation that spans across linearly coupled spatial modes.
% Figure of three modulation formats
\begin{figure*}[!bht]
% \psfrag{SBS}[c][c][1][0]{SBSREE}
\centering
\includegraphics[width=1.8\columnwidth]{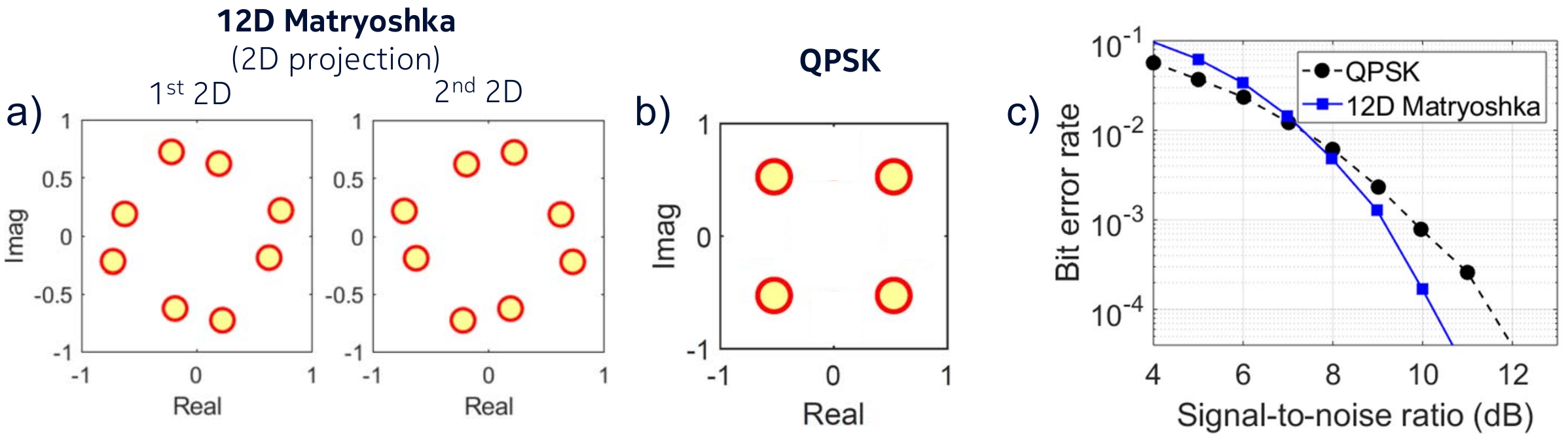}
\caption{a) 2D projections of the proposed 12D Matryoshka modulation format along with b) the (2D) QPSK format. The three pairs of 2D projections of the 12D Matryoshka format are identical to the pair shown. Selection rules between the six 2D sets of the 12D format generate $2^{12}$ points; c) BER curves for both formats.}
\label{fig:12Dformats}
\end{figure*}

\section{12D Matryoshka Modulation Format}
 \label{sec:12Dformat}

The design of multidimensional modulation formats is challenging in a few different ways as there are competing goals to provide increased transmission performance. An important objective is to achieve high SE with good immunity to AWGN while keeping the constellation complexity low. The latter helps for bit mapping of constellations and can facilitate hardware implementation. In addition to these properties, for fiber-optic transmission, one wants these constellations to generate minimum nonlinear distortions, which often put additional constraints on the design of the format.

Figure~\ref{fig:12Dformats}a displays two 2D projections of the first four dimensions of the proposed 12D modulation format, referred to as {\it 12D Matryoshka}. The other two pairs of two 2D projections that complete the six 2D projections are omitted here as they are identical to the first two shown. All six 2D projections along with selection rules constitute the 12D Matryoshka format. Selections rules between points across 2D sets reduce the unconstrained $8^6 = 2^{18}$ constellation points to $2^{12}$ points. One notices that each 2D projection of the 12D format has two sets of four points on two (nearly identical) radii offset by an angle of 0.15 radians. We impose a rule of selecting only points on the different radii within the three pairs of 2D sets that form a 12D format. The format is then bit mapped using Gray code for each of the two rings in each 2D sets with a single bit added for ring selection. All 12D constellation points having either even or odd bit parity are then removed (constellation puncturing), leaving 4096 points in 12D. The resulting Matryoshka constellation is then bit mapped again with 12-bit sequences. This results in a SE of 4~bits/s/Hz for each wavelength, identical to the PDM-QPSK format shown in Fig.~\ref{fig:12Dformats}b. The different sizes of the constellation points between the two formats reflect their frequency of use. The 12D Matryoshka format has constant power in 4D, and therefore in 12D, as for PDM-QPSK. Figure~\ref{fig:12Dformats}c displays the BER curves for both formats showing better BER for the proposed Matryoshka format at an SNR > 7.3~dB, or a BER < $9 \times 10^{-3}$.

%
%Note that the 2D projection of the 12D format shows that it is constant power in 2D, and consequently in 4D and 12D. This is the same properties as the QPSK format that helps provide excellent nonlinear transmission for this format. Another good property of QPSK is that it is easily Gray bit mapped providing a single bit difference to neighboring constellation points. Figure~\ref{fig:BER12DvsQPSK} shows the BER curves of the 12D and QPSK formats showing benefit of the multidimensional format proposed at SNR > 7.3~dB (BER < 9 x $10^{-3}$). The bit map was based on \
%% Figure of three modulation formats
%\begin{figure*}[!tbh]
%% \psfrag{SBS}[c][c][1][0]{SBSREE}
%\centering
%\includegraphics[width=0.8\columnwidth]{Fig_Plot_12D_BonSE4_vs_QPSK}
%\caption{Bit-error rates of the proposed 12D format and QPSK in AWGN.}
%\label{fig:BER12DvsQPSK}
%\end{figure*}

%\begin{figure}[b]
%\centering\includegraphics{fig1.eps}
%\caption{(a) Sample graph with pink (dotted), blue (solid) and black (dashed) lines, (b) Sample graph with black (diamond), black (triangle) and black (square) markers.}
%\label{fig1}
%\end{figure}
% Figure of set-up
\begin{figure*}[!bht]
\centering
\includegraphics[width=1.99\columnwidth]{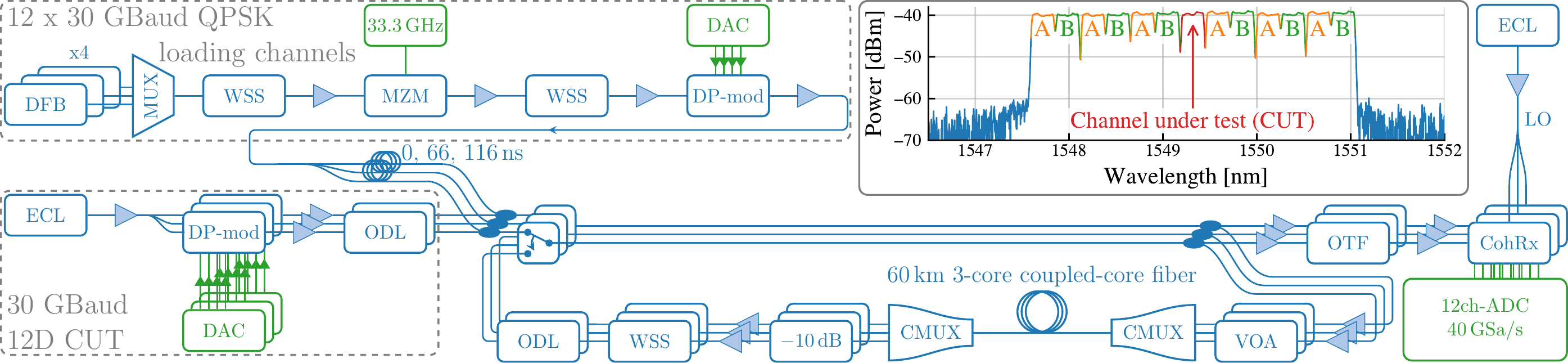}
\caption{Experimental transmission set-up.}
\label{fig:setup}
\end{figure*}

\section{Experimental Set-Up}
  \label{sec:experiment}
% 5th-order DM-MZM ? ASK ROLAND
The experimental setup shown in Fig.~\ref{fig:setup} consists of a transmitter producing 13 wavelength-division multiplexed (WDM) channels spaced at 33.3~GHz and modulated at 30~Gbaud. Three time-synchronized four-channel 60~GSa/s digital-to-analog converters (DACs) were used to create the 30~GBaud 12D Matryoshka modulation format described in Sec.~\ref{sec:12Dformat} as the channel under test (CUT). It is inserted in the middle of the signal spectrum (see inset in Fig~\ref{fig:setup}). The DACs drive three dual-polarisation modulators (DP-mods), based on dual-polarisation double-nested Mach-Zehnder modulators (MZMs), each modulating a copy of a tone at 1549.3~nm provided by an external cavity laser (ECL). The three tributaries of the 12D signal are time-aligned with an accuracy of $\pm 4$~ps using two optical delay lines (ODLs) placed in the second and third tributary path. The twelve 30~GBaud PDM-QPSK  channels  are  created  by modulating six tones of an optical frequency comb. A 120~GSa/s DAC provides the driving signals to a fourth DP-mod, creating two uncorrelated  PDM-QPSK channels out of each tone. The 12D CUT and the loading channels are combined and injected into a recirculating  loop. The three spatial tributaries are launched into a 60~km spool of three-core coupled-core multicore fibre (CC-MCF)~\cite{ryf2012c} using a core multiplexer (CMUX), consisting in a tapered structure similar to a photonic lantern that  couples  three  input fibers into the three cores of the CC-MCF, resulting in an added insertion loss of $(0.7 \pm 0.2)$~dB. Since the recirculating loop components are single mode,  a  second  CMUX  is used to separate the output of the CC-MCF into three tributaries. They are subsequently amplified, filtered, and launched into the CC-MCF again, where variable optical attenuators (VOAs) where used to change the launch power into the fiber cores and wavelength-selective switches (WSSs) were optimized to correct for the launch power dependent spectral tilt. The ODLs insured that the three tributaries experienced identical delays, within the accuracy of $\pm 4$~ps.
The three tributaries are extracted from the recirculating loop by using splitters with a splitting ratio of 10:90 followed by  optical amplifiers an optical tunable filters (OTF) to select the CUT, followed by three polarisation-diverse intradyne coherent receivers (cohRxs) connected to a single local oscillator (LO) and a 12-channel 40~GSa/s digital storage oscilloscope. The captured signals are processed offline by first up-sampling the signals to 2 samples/symbol, performing  chromatic dispersion and frequency-offset compensation followed by timing identification and a $6 \times 6$ MIMO processing, based on a frequency domain equalizer with 1000~symbol-spaced taps, followed by a phase recovery for each spatial tributary. The reconstructed fields are subsequently processed as 12D fields and SNR, MI and GMI are calculated.
The 60-km long CC-MCF~\cite{ryf2012c} had an effective area of 120~$\mu$m$^2$, and a loss of 0.18~dB/km, and an additional 10~dB attenuation was added to increase the span loss to $(22.8 \pm 0.2)$~dB, resulting in a loss equivalent span length of 115~km. The additional attenuation was necessary to produce lower SNR values at fewer loop recirculation numbers since starting from $\sim$60~loops recirculations, the measurements became more susceptible to loop artifacts and the effects of an increased in mode-dependent loss (MDL) were observed.
% Figure of MI and GMI
\begin{figure*}[!tbh]
\centering
\includegraphics[width=1.95\columnwidth]{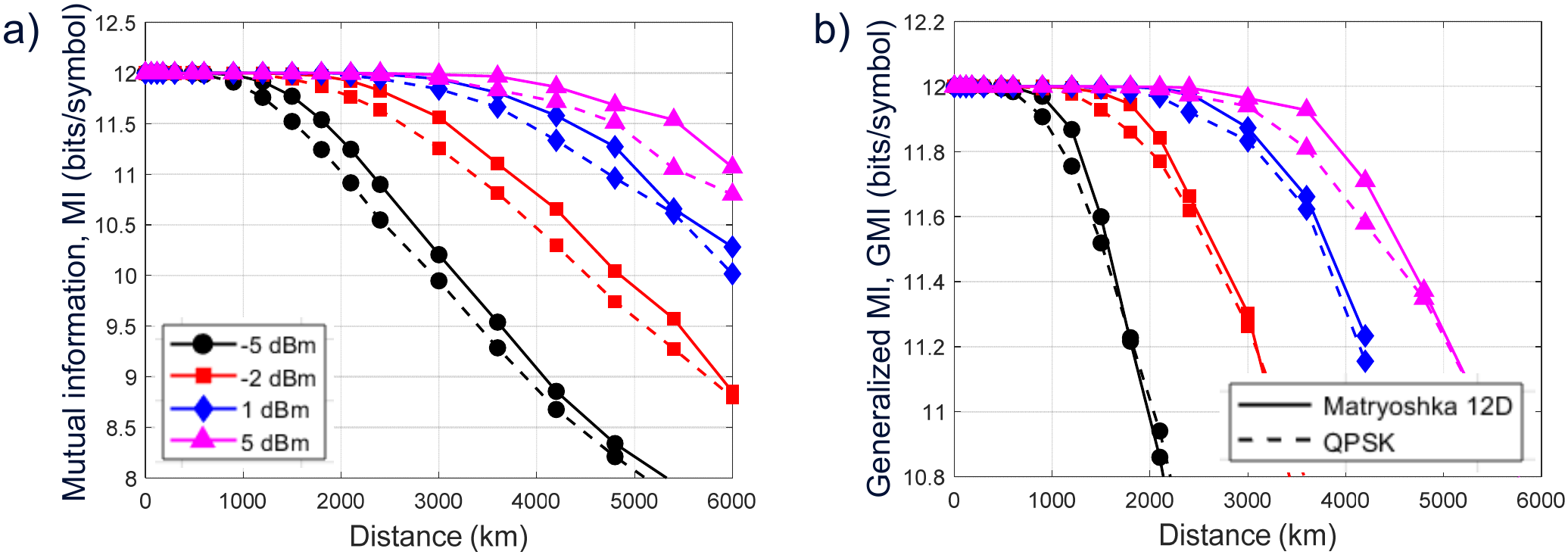}
\caption{a) Mutual information (MI) as a function of distance for the 12D Matryoshka and PDM-QPSK for different powers per channel. The MI of PDM-QPSK is reported over 12D here for a fair comparison; b) Generalized mutual information (GMI) for the same power levels.}
\label{fig:MIvsDistance}
\end{figure*}

\section{Transmission}
  \label{sec:transperf}
The mutual information (MI) after transmission of the two modulation formats are shown in Fig.~\ref{fig:MIvsDistance}a. The MI of PDM-QPSK is reported over the entire set of three spatial modes (i.e. 12D) to allow a fair comparison between the two formats. Both modulation formats can carry 12 bits of information after a few spans. After 10 to 20 spans depending on the power, the MIs of the two formats start to differ with the 12D Matryoshka outperforming PDM-QPSK at all distances and powers. At an MI = 11~bits/symbol, the 12D Matryoshka format allows an averaged increase of 8~to~15\% of the propagation distance. There appears to be a slight increase in the difference of MIs at high powers favoring 12D Matryoshka over PDM-QPSK. Further investigation is needed for a more precise assessment in this regime and in the presence of MDL.
% Figure of SNR
\begin{figure}[!tbh]
\centering
\includegraphics[width=0.85\columnwidth]{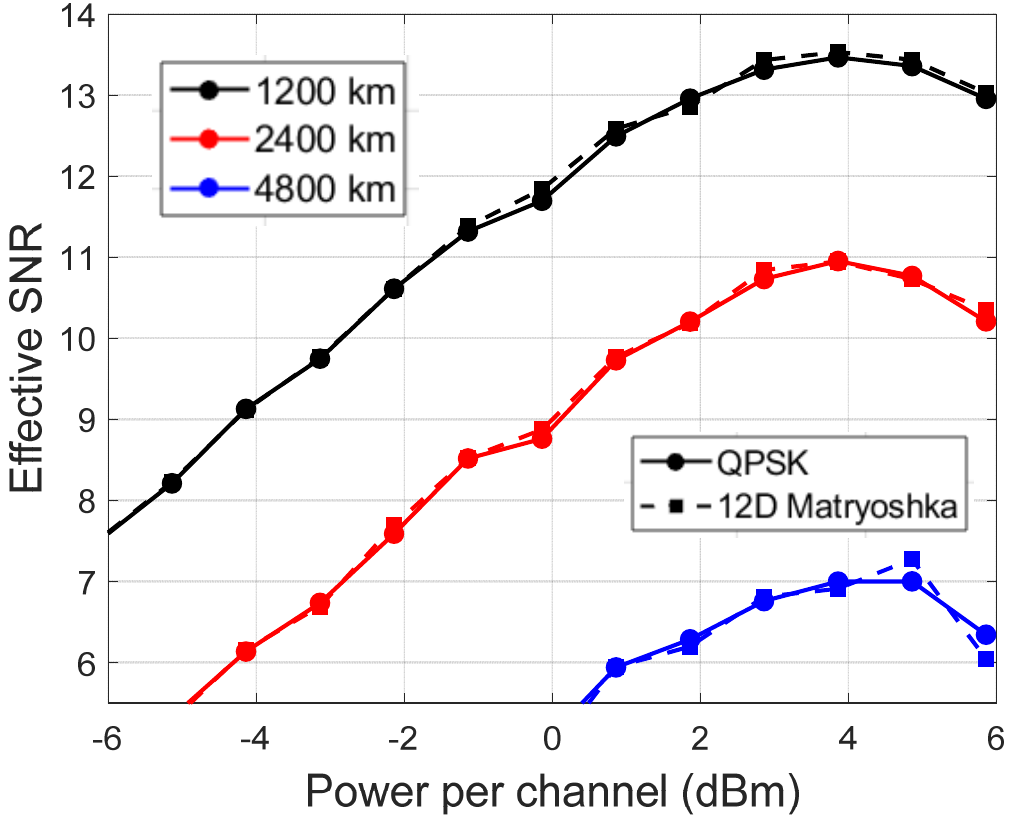}
\caption{Effective SNR as a function of signal power for 1,200, 2,400 and 4,800~km for the 12D Matryoshka and PDM-QPSK.}
\label{fig:SNRvsDistance}
\end{figure}

The generalized MI (GMI) for the same conditions as the MI measurements of Fig.~\ref{fig:MIvsDistance}a is reported in Fig.~\ref{fig:MIvsDistance}b. It shows that most of the benefits of the 12D format for MI are preserved at high values of GMI, including in the high power regime. The bit mapping therefore appears to offer `resistance' to the types of distortions produced during transmission.
% The benefit of the 12D constellation decreases to eventually vanish or be negative for lower values of GMI.

Figure~\ref{fig:SNRvsDistance} shows the effective SNR as a function of power at a few different transmission distances. The effective SNR is calculated using the variances of the constellation points. The SNRs based on this variance are virtually identical for the two modulation formats with both exhibiting a maximum around 4~dBm per channel. Note that the variance is insufficient to describe the complete noise distribution.

\vspace{-1mm}
\section{Conclusion}
 \label{sec:conclusion}
We presented the first long-distance transmission of a 12D modulation format implemented across coupled spatial modes in a 3-core coupled-core multicore fiber. The format shows improved performance over PDM-QPSK, both having spectral efficiencies of 4~bits/s/Hz.

% \section{Acknowledgment}
%  \label{sec:ack}
We would like to acknowledge R. W. Tkach, A. Ghazisaeidi, G. Gavioli and S. Weisser for helpful discussions.
%\section{References}
% \bibliographystyle{IEEEtran}
% \bibliography{C:/work_le1/Biblio/BibTeX/Essiambre_References_Database_ver104}
% \bibliography{D:/work_ti1/Biblio/BibTeX/Essiambre_References_Database_ver104}
%\section{References}
%References may use a fourth page. This additional page is for references \underline{only}. Footnotes, should appear as footnotes, and should not be included within the references section. Please ensure you follow the referencing style guide on the next page. If the style guide is not used the paper may be rejected.

\clearpage

% Generate Biblio with this
% \bibliographystyle{IEEEtran}
% \bibliography{C:/work_le1/Biblio/BibTeX/Essiambre_References_Database_ver105}
% \bibliography{D:/work_ti1/Biblio/BibTeX/Essiambre_References_Database_ver105}

\bibliographystyle{abbrv}
\end{document}